\pdfoutput=1 

\documentclass{pnastwo-anon}
\usepackage{graphics,amssymb,amsmath,bm}

\begin{document}

\newcommand{\SOM}{appendix}

\renewcommand{\ee}{\mathrm{e}}
\newcommand{\tobs}{t_\mathrm{obs}}
\newcommand{\eps}{\epsilon}
\newcommand{\sig}{\sigma}
\newcommand{\ii}{\mathrm{i}}
\newcommand{\CC}{\mathcal{C}}
\newcommand{\WW}{\mathbb{W}}
\newcommand{\HH}{\mathbb{H}}
\newcommand{\EE}{\mathbb{E}}
\newcommand{\FF}{\mathcal{F}}

\newcommand{\rate}{\lambda}
\newcommand{\upr}{\gamma}

\newcommand{\zz}{\ee^{-s}}
\newcommand{\zzs}{\ee^{-s^*}}

\newcommand{\Sc}{S}

\title{Finite-temperature critical point of a glass transition}
\author{Yael S. Elmatad\affil{1}{Department of Chemistry, University of California, Berkeley, CA 94720, USA},
  Robert L. Jack\affil{2}{Department of Physics, University of Bath, Bath BA2 7AY, United Kingdom},
  Juan P. Garrahan\affil{3}{Department of Physics and Astronomy, University of Nottingham, 
    Nottingham, NG7 2RD, United Kingdom},
  and
  David Chandler\affil{1}{Department of Chemistry, University of California, Berkeley, CA 94720, USA} 
}



\maketitle

\begin{article}  

\begin{abstract}
We generalize the simplest kinetically constrained model of a glass-forming liquid by softening kinetic constraints, allowing them to be violated with a small finite rate.  We demonstrate that this model supports a first-order dynamical (space-time) phase transition, similar to those observed with hard constraints.  In addition, we find that the first-order phase boundary in this softened model ends in a finite-temperature dynamical critical point, which we expect to be present in natural systems.  We discuss links between this critical point and quantum phase transitions, showing that dynamical phase transitions in $d$ dimensions map to quantum transitions in the same dimension, and hence to classical thermodynamic phase transitions in $d+1$ dimensions.  We make these links explicit through exact mappings between master operators, transfer matrices, and
Hamiltonians for quantum spin chains.
\end{abstract}


\abbreviations{
KCM,~kinetically constrained model; 
FA,~Fredrickson-Andersen; 
sFA,~`softened' FA;
QPT,~quantum phase transition
}


\section{Introduction}

As a liquid is cooled through its glass transition, it freezes
into an amorphous solid state, known as a glass~\cite{glass}.  The transition from
fluid to solid typically requires only a small change in temperature
and is accompanied by characteristic large fluctuations, known
as dynamical heterogeneity~\cite{DH}.  Based on these observations, several 
theories draw analogies between the glass transition and
phase transitions that occur in model systems~\cite{RFOT,Tarjus-Kiv,mct,GC-review}.  
However, experimental glass transitions are not thermodynamic
phase transitions, since responses to thermodynamic fields such
as temperature or pressure 
are never observed to 
diverge.  On the other hand, models of glass-forming liquids have been shown to exhibit a dynamical phase transition, which is controlled not just by temperature, but also by a biasing field that moves the system away from equilibrium~\cite{Merolle,spacetime-jcp,
Garrahan-Fred-PRL,Garrahan-Fred-JPA, Hedges-science,s-ROM}.  
Here, we demonstrate the existence of a non-trivial critical point associated with this transition.

The phase transition that we consider occurs in trajectory space.  
We bias the system towards an `ideal glass' phase by enhancing the probability of trajectories where 
particle motion is slow.  The parameter that controls this bias -- the field $s$ -- can be varied continuously in computer simulations.  For models of glass-forming liquids, the response to this change can be large and discontinuous, corresponding to a first-order phase transition.  Such transitions between an ergodic (fluid) state and a non-ergodic (glass) state were first demonstrated~\cite{Garrahan-Fred-PRL} for kinetically-constrained lattice models (KCMs)~\cite{Ritort-Sollich}.  More recently, evidence for such a transition has been found in the molecular dynamics of an atomistic model of a supercooled liquid~\cite{Hedges-science}.  

In KCMs, the transition to the inactive phase takes place when the biasing field $s$ is infinitesimally small,
and this result is independent of the temperature.  However, very general arguments based on ergodicity breaking (see, for example, Refs.~\cite{s-ROM,Gaveau-Schulman,BK}) indicate that this transition must disappear at high temperature in molecular systems, and it is also expected
that transitions in ergodic molecular fluids should take place at finite 
$s$.  
In this context, the essential difference between KCMs and molecular systems is that forces constraining dynamics in the former are infinite (i.e., hard) while those in the latter are finite (i.e., soft).  Infinitely long-lived inactive metastable states exist in KCMs because the constraints are hard.   The dynamic first-order phase transition detailed in Refs.~\cite{Garrahan-Fred-PRL,Garrahan-Fred-JPA} coincides with a non-equilibrium biasing that stabilizes these inactive states.  But because constraints in molecular systems are soft, it is not obvious that this first-order transition will persist in natural systems.  Here, we address this issue by considering the effects of softening constraints in the simplest of all KCMs -- the one-spin facilitated Fredrickson-Andersen (FA)~\cite{FA} model.

In this softened FA (sFA) model, we find two main results, illustrated schematically in Fig 1. Firstly, we prove that the dynamic phase transition found in the original FA model still occurs when the constraint forces are finite, but the transition now takes place at non-zero $s$. This result means that the sFA model -- a system with finite short-ranged forces of interaction, one that exhibits long relaxation times and dynamic heterogeneity without equilibrium thermodynamic transitions of mode-coupling theory~\cite{mct} and random first-order theory~\cite{RFOT} -- has a dynamical non-equilibrium glass transition.  The solution of the sFA model is therefore a concrete, if overly simplified, illustration of the picture of the glass transition proposed in Refs.~\cite{GC-PRL02,Merolle} and later supported by the results of 
Refs.~\cite{spacetime-jcp,Garrahan-Fred-PRL,Garrahan-Fred-JPA,Hedges-science}. 
We believe that this is the first study to demonstrates all of these features in a single model.

Our second main result is that the sFA model supports a new finite-temperature critical point
that only appears when the constraints are softened.  
More specifically, we show that the first-order dynamical phase boundary 
in the sFA model ends at a critical
point, with universal scaling behavior that maps to that of a quantum-Ising model 
in a transverse field~\cite{Sachdev}, and hence
to a classical liquid-vapor (or liquid-liquid) critical point. 
The existence of such a critical point
is consistent with the requirement that models of glass-formers should recover
simple liquid behavior at high temperatures.

\begin{figure}
\begin{center}
\resizebox{3.5cm}{!}{\includegraphics{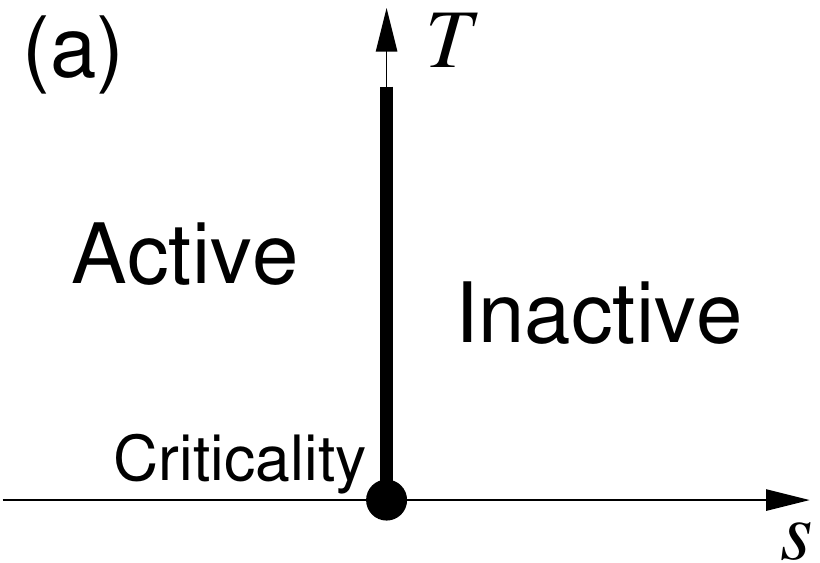}}
\hspace{24pt}
\resizebox{3.5cm}{!}{\includegraphics{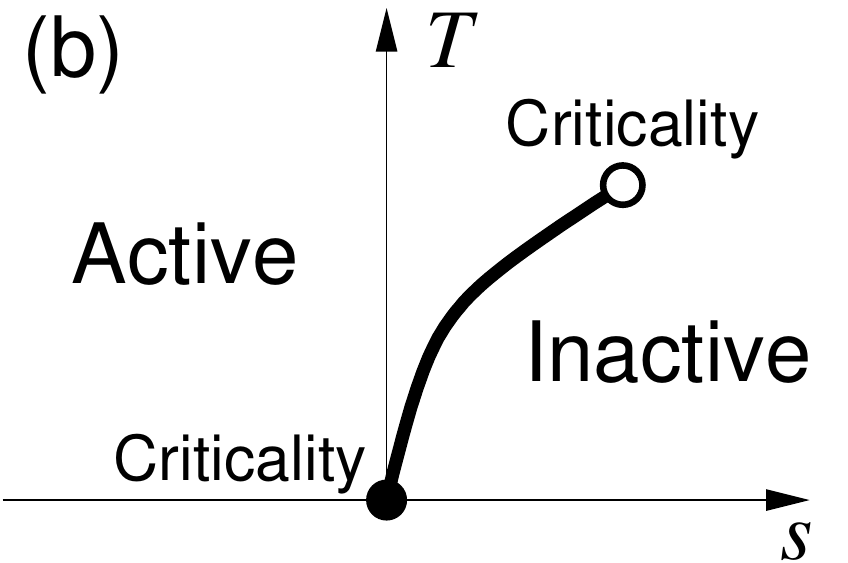}}
\end{center}
\caption{ \label{fig:phase_sketch}
{\bf(a)}~Generic space-time phase diagram for KCMs~\cite{Garrahan-Fred-PRL}. 
There is a first order phase boundary that occupies the $s=0$ axis, separating
an active fluid phase from an inactive `glass'.  The critical point $s=T=0$ is identified
with a filled circle: no motion takes place in 
this state, and the approach to this point is characterized
by scaling behavior and slow
dynamical relaxation~\cite{Whitelam,FA-AA}.
{\bf(b)}~Sketch of the space-time phase diagram for the softened FA model, under
the assumption that the probability of violating constraints $\eps$ has an Arrhenius
form,
as described in the main text.
The first-order phase boundary moves away from the $s=0$ axis and ends
in a new finite-temperature critical point, identified with an open circle.
The scaling behavior in the vicinity of this
point is analogous to the critical
behavior near liquid-vapor 
transitions, and is
different from the scaling near $s=T=0$.
} 
\end{figure}

\section{Softened FA model}

Dynamics in supercooled liquids is heterogeneous~\cite{DH}: 
particle motion is relatively significant in some regions of space, and relatively insignificant in others.
We associate mobile regions with `excitations' that facilitate
local motion. Near such excitations, motion takes place
through fast processes with rate $\rate$, while motion in immobile 
regions takes place with a slower rate $\rate\epsilon$.  To arrive at KCMs,
one sets $\epsilon=0$, for which the physical picture is that
relaxation in the system is dominated by correlated 
sequences of fast processes and that the slow processes are irrelevant~\cite{GC-PRL02,GC-pnas}.  

The sFA model consists of 
a set of binary variables (spins) $n_i=0,1$, where $n_i=1$
denotes the presence of an excitation. 
The rate for flipping spin $i$ from $1\to0$ is
$[r_i]_{1\to0}=\rate C_i$,
where $C_i$ is a constraint function 
that depends on the neighbors of spin $i$, taking a value of order
$\epsilon$ if the site is in a slow region, and a value of order unity
if the site is in a fast region.  The reverse process, $
0\to1$, takes
place with a slower rate $[r_i]_{0\to1}=\upr\rate C_i$. 
To ensure detailed balance at a temperature $T$, 
we take $\upr = \exp(-J/T) = c/(1-c)$, where $J$ is the energy associated with creating an excitation, and $c = \langle n_i \rangle$ is the equilibrium average concentration of excitations. 
(We use units of temperature such that Boltzmann's constant $k_\mathrm{B}=1$.)
The case $\eps = 0$ would be a KCM, while we refer to models with $\eps > 0$ as `softened' KCMs. We expect that violating a kinetic constraint should require a large activation energy $U > J$, so that  $\eps \propto \exp(-U/T) $.

In the sFA model, we take $C_i =\sum_{j\in \mathrm{nn}(i)} [n_j + (\epsilon/2)]$, where the sum runs over the nearest neighbors $j$ of site $i$. We also allow excitations to hop (diffuse) between adjacent sites: the process where the state $(1,0)$ of two nearest neighbors changes to $(0,1)$ occurs with rate $\rate D$. The value of the (dimensionless) diffusion constant $D$ has no qualitative effect on the behavior of the model, but it makes some of our later calculations more tractable 
because it allows us to 
solve analytically for the position of the critical point shown in Fig.~\ref{fig:phase_sketch}(b). 
(This technical aspect is detailed below and in the appendix.) 
We fix the units of time by setting  $\rate = 1$, so the state of the sFA model is specified by the three dimensionless parameters $(\upr, \epsilon, D)$.  The one-spin facilitated FA model is recovered by setting $\epsilon  = D = 0$. In constructing Fig. 1, we assumed that $\upr$ and $\epsilon$  both depend on temperature as described above,
and we take $U > 3J$. 
so that the model behaves as a KCM in the limit where $T \rightarrow 0$.

Finally, the methods that we use require us to specify
an observable or order parameter, $K$, that measures the amount of dynamical
activity in a trajectory (or history) of the model.  A suitable quantity must be extensive in the volume of space-time. 
We work in continuous time where successive
configurations differ by exactly one local change (diffusion of an excitation or flipping
of a spin).  We define the dynamical activity $K$~\cite{Merolle,Garrahan-Fred-PRL,Fred-JSP,
Hooy,Maes} to be the total number of such configuration changes in a trajectory.

\section{Active and inactive space-time phases}

Fig.~\ref{fig:phase_sketch} shows two space-time phase diagrams.
The idea of a space-time phase is at the heart of the work presented here, and we now
explain this notion in more detail.  In statistical mechanics,
a `phase' (such as a liquid or crystal) 
is a region of phase space in which configurations are macroscopically homogeneous
and share similar qualitative
properties.  By analogy, a `space-time phase' is a region of trajectory space (a set of trajectories)
with similar qualitative features.  
 
\begin{figure*}
\begin{center}
\resizebox{11.8cm}{!}{\includegraphics{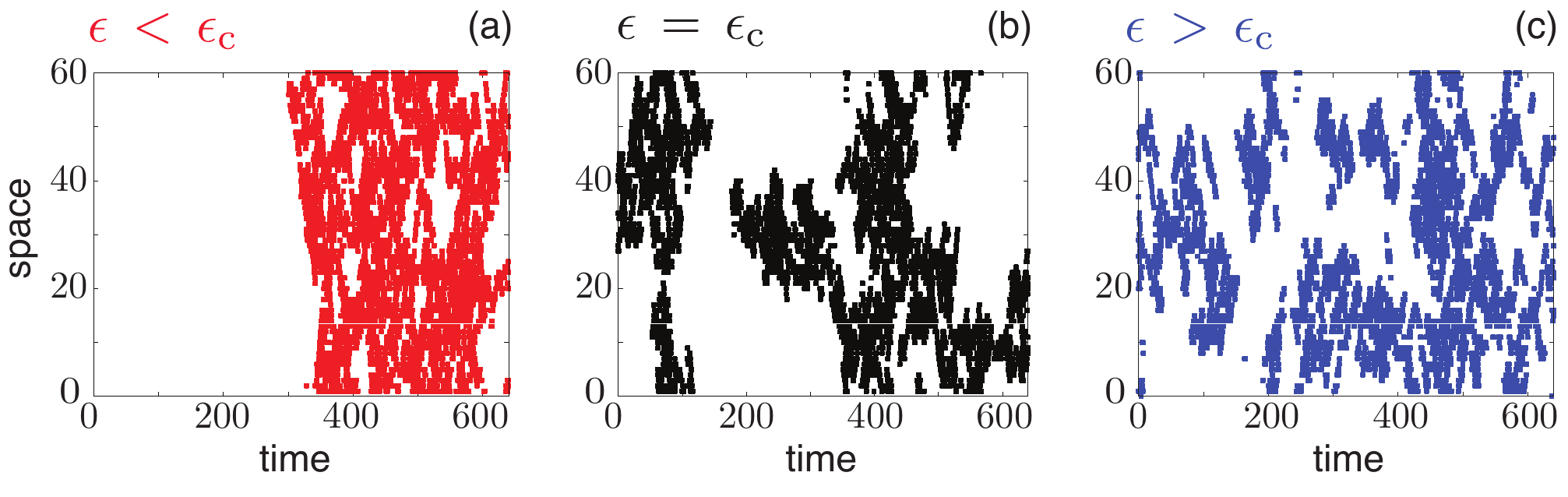}}
\caption{\label{fig:trajs} Trajectories obtained by 
constraining the activity $K$, for the three state points identified in Fig.~\ref{fig:phaseD}. 
Active sites ($n_i=1$) are colored,
with inactive sites ($n_i=0$) are white.
{\bf(a)} For small $\eps$, we observe phase separation
in time.  {\bf(b)} At a specific value $\eps=\eps_\mathrm{c}$, the system is critical, with
large self-similar clusters of active sites. {\bf(c)} For large $\eps$, the system exhibits only
one phase, and all correlations have finite range.
}
\end{center}
\end{figure*}

Two space-time phases are depicted in Fig.~\ref{fig:trajs}(a).  Using computational
methods that we will discuss below, we obtained many trajectories for the sFA model, 
covering a wide range of $K$.
Working always with the same values of $(\upr,\eps,D)$, 
we then restrict ourselves to trajectories where the value of $K$
is equal to approximately half of its equilibrium average.  By analogy with equilibrium
statistical mechanics, we refer to the set of trajectories with a given value of $K$ as
a `microcanonical' ensemble of trajectories.  For the parameters of
 Fig.~\ref{fig:trajs}(a), this restriction leads to
``phase separation in time'': the trajectory that we show is representative
of the ensemble, and it has an early part that
is inactive (very few spin flips), while its later part is much more active (many spin flips).  
(The ensemble is time-reversal
symmetric, so one might similarly have observed a trajectory where the early part is active
and the later part inactive.)

In Ref.~\cite{Garrahan-Fred-PRL}, 
it was proven that phase separation in time occurs for the FA model (i.e., the model with $\eps=0$).  
The main subject of our current article is how this behavior is affected by a finite
value for $\eps$, since this relaxes the assumption of infinite forces of constraint
that was used in~\cite{Garrahan-Fred-PRL}.
The three panels of Fig.~\ref{fig:trajs} summarize these effects.  
Phase separation in time is clear in Fig.~\ref{fig:trajs}(a), even though $\eps$ is non-zero.
On the other hand, there is no such   
effect in Fig.~\ref{fig:trajs}(c): whatever restriction we place on $K$,
phase separation never occurs for the values of $(\upr,\eps,D)$ used that figure.
The qualitatively different situations shown in Fig.~\ref{fig:trajs}(a) and 
Fig.~\ref{fig:trajs}(c) are separated by a critical point in space-time: a representative trajectory
from the critical system is shown in Fig.~\ref{fig:trajs}(b).  
There are active and inactive
domains with a range of sizes, and the interfaces between them are diffuse and complex. 

\subsection{Biased ensembles of trajectories}

We introduce a biasing field $s$ which couples to $K$ in the same
way that the inverse temperature couples to the energy in the canonical ensemble of equilibrium statistical mechanics.
In so doing, we define non-equilibrium ensembles of trajectories~\cite{Merolle,spacetime-jcp,Garrahan-Fred-PRL,Ruelle,Fred-JSP} known
as $s$-ensembles, that can be studied both analytically and computationally.  A detailed description of the application of our methods to KCMs is given in Ref.~\cite{Garrahan-Fred-JPA}.  In the $s$-ensemble, $\langle A \rangle_s$ denotes the mean value of a function of system history, $A$.  Denoting unbiased equilibrium averages by $\langle A \rangle_0$, we have
\begin{equation}
\langle A \rangle_s = \langle A \ee^{-sK} \rangle_0 \frac{1}{Z(s,\tobs)},
\label{equ:s-vanilla}
\end{equation}
where 
$ Z(s,\tobs) = \langle \exp(-sK) \rangle_0 $
is the partition function for the $s$-ensemble.  The ensemble
with $s=0$ is simply the unbiased equilibrium dynamics of
the sFA model.
We define a dynamical
free energy (per unit time):
\begin{equation}
\psi(s)=\lim_{\tobs\to\infty}\frac{1}{\tobs} \log Z(s,\tobs).
\label{equ:def_psi}
\end{equation}
With these definitions, 
microcanonical (fixed-$K$) and canonical (fixed-$s$) ensembles are equivalent, in the same
sense as equivalence of ensembles in statistical mechanics. 
Thus, while the field $s$ does not have a direct physical interpretation,
it plays the same role as a constraint on the activity $K$.

Applying this formalism to the sFA model, we arrive at
ensembles of trajectories that depend on four parameters $(\upr,\eps,D,s)$.  
Within these ensembles, the sFA model has no conserved or constrained quantities, so
phase separation in time is no longer possible.  
Instead, one finds phase coexistence: for a particular field $s=s^*$,
active and inactive space-time phases have equal free energies, while the inactive phase is found for $s>s^*$
and the active one for $s<s^*$.  
A useful result that we derive in the \SOM~is that active and inactive space-time phases in the sFA
model are related by a symmetry transformation.  This symmetry means that space-time phase
coexistence can occur only when
\begin{equation}
\frac{1+\upr}{1+\eps} = 
              \sqrt{[1-\upr-D(1-\ee^{-s^*})]^2 + 4\ee^{-2s^*}\upr}
                      - (1-\ee^{-s^*})D
\label{equ:dual}
\end{equation}
For an
sFA model 
specified by $(\upr,\eps,D)$, there is at most one solution for $s^*$.
If space-time phase coexistence occurs for 
parameters, then the coexistence point is $s=s^*$.  

\subsection{Computational sampling of space-time phases} 

\begin{figure}
\begin{center}
\resizebox{6.8cm}{!}{\includegraphics{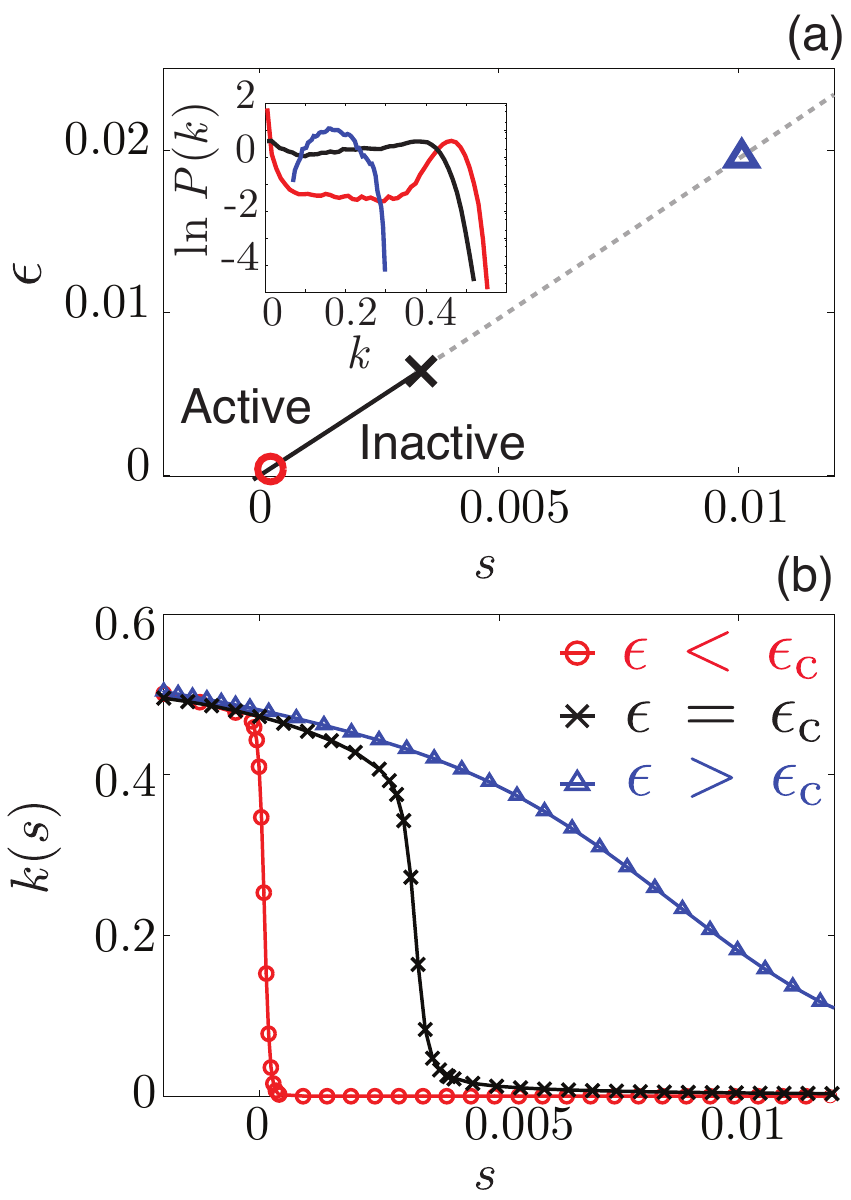}}
\end{center}
\caption{\label{fig:phaseD} 
{\bf(a)} Phase diagram of the $1d$ sFA model. For $\upr=0.25$, we show the $(s,\eps)$ plane, varying the diffusion constant $D$ as a function
of $\eps$, as discussed
in the text.  The solid line is a first-order phase boundary between active and inactive states.
It ends at a critical point.  The dashed line shows the extension of the symmetry 
line~(Equ. \ref{equ:dual}) into the one phase region. The symbols show the parameters for which we present
data, which are $\eps=1.9\times10^{-4}, 0.0063, 0.01$.  
{\bf(Inset)} Histograms of the (intensive) activity $k=K/(N\tobs)$, in symmetrized $s$-ensembles
corresponding to the three symbols in the main figure. 
{\bf(b)} Plot of the activity $k(s)$, at the state points identified in (a).} 
\end{figure}

\begin{figure}
\begin{center}
\resizebox{7.4cm}{!}{\includegraphics{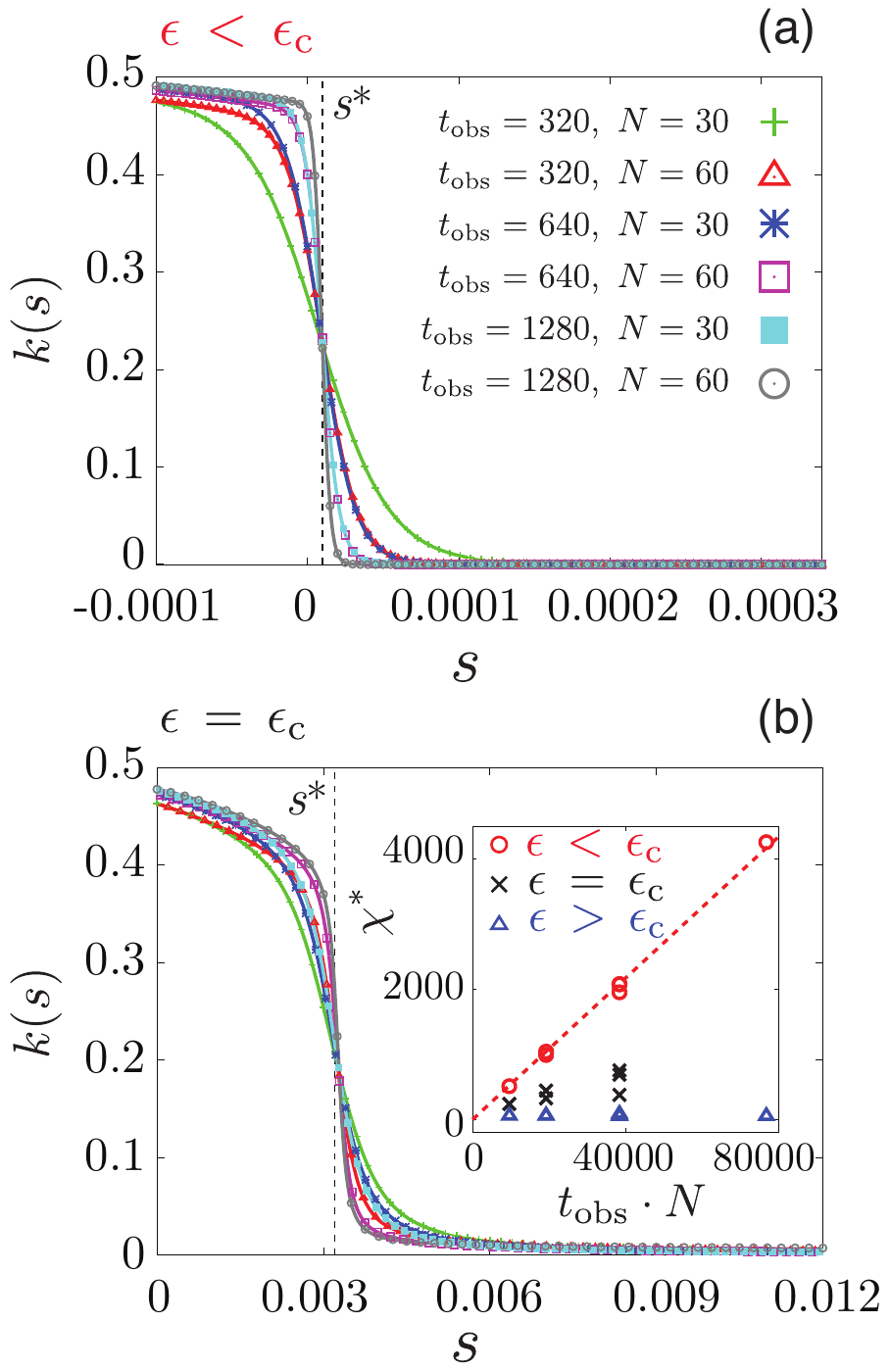}}
\caption{\label{fig:scaling} 
Finite-size scaling in the $s$-ensemble. We show the dependence of the mean activity
on the biasing field $s$, for various system sizes and observation times.  Dashed vertical
lines show the positions of first-order and second-order transitions predicted by Equ.~\ref{equ:dual}.  
{\bf (a)} Data for 
$\eps<\eps_\mathrm{c}$, at the state point marked by a circle in Fig.~\ref{fig:phaseD}(a).  {\bf (b)} Data
for $\eps=\eps_\mathrm{c}$, for the state point marked by a cross ($\times$) in Fig.~\ref{fig:phaseD}(a).  
The values of $N$ and $\tobs$ are the same in panels (a) and (b).  {\bf Inset to (b)}: The derivative
of the activity with respect of the field $s$, evaluated at $s=s^*$, for the three state
points shown in Fig.~\ref{fig:phaseD}(a), varying $N$ and $\tobs$.
}
\end{center}
\end{figure}

We have investigated
the $s$-ensemble computationally, using transition path sampling (TPS)~\cite{TPS}.  
The procedure is essentially the same as that used in Ref.~\cite{Hedges-science}: 
we run standard
TPS simulations with shooting and shifting moves, and a Metropolis acceptance
criterion based on values of $s$ and $K$.  
As discussed above, the behavior of the sFA model
does depend on the parameter $D$, but qualitative features are largely independent of $D$.  
We have verified this by performing numerical simulations for several values of $D$, including $D=0$.  
However, in this article,
we exploit this adjustable parameter to our advantage.  
Specifically, in Figs.~\ref{fig:trajs}-\ref{fig:scaling}, we fix $\upr$ and vary $\eps$, 
taking 
\begin{equation}
D = \textstyle{\frac12} [ 1 - \upr + \sqrt{ (1-\upr)^2 + 4\zzs\upr} ].
\label{equ:freef}
\end{equation}
where the value of $s^*$ is obtained by solving Equ.~\ref{equ:dual} simultaneously with
Equ.~\ref{equ:freef}.  
With $D$ constrained in this way, and the coexistence field $s^*$ being known, it becomes
possible to make considerable analytic progress with the sFA model, as we discuss
in the \SOM.  In particular, we can then solve for the value of the 
parameter $\eps=\eps_\mathrm{c}$ at which the system becomes critical.  This solution
 was used to generate Fig.~\ref{fig:trajs}(b).

As in standard Monte Carlo simulations near phase coexistence,
 good sampling of the $s$-ensemble may be frustrated by large free energy
barriers between coexisting phases.  To avoid this, we work
at $s=s^*$ and ensure that the system explores both active and inactive
phases within each TPS run~\cite{Hedges-science}.
Other obstacles to accurate characterization of phase coexistence
include the possibility of large boundary effects.
While periodic boundary conditions are used for the spatial degrees
of freedom in the sFA model,
it is not possible to use periodic boundaries in time within our computational approach
(for an analytic treatment, see 
Ref.~\cite{BK}).  
If $s>0$ then this means the initial and final parts of the trajectory are biased
towards the active phase~\cite{Garrahan-Fred-JPA}.  To reduce this effect and more accurately
characterize phase coexistence in the sFA model,
we introduce a refinement to the method of 
Ref.~\cite{Hedges-science}.  We bias the 
initial and final conditions in our ensembles of trajectories, arriving at
a `symmetrized $s$-ensemble' that fully respects the symmetry between
active and inactive phases in the sFA model (see \SOM). 
Expectation values in this ensemble are given by
\begin{equation}
\langle A \rangle_{s,\mathrm{sym}} 
=
\langle A \ee^{-sK + g[\mathcal{N}(0)+\mathcal{N}(\tobs)]} \rangle_0 \frac{1}{Z_\mathrm{sym}(s,\tobs)}
\label{equ:ens-sym}
\end{equation}
with ${Z_\mathrm{sym}(s,\tobs)} = \langle \exp\{-sK + g[\mathcal{N}(0)+\mathcal{N}(\tobs)]\} \rangle_0$.
Here $\mathcal{N}(t)=\sum_i n_i(t)$ is the total number of excitations in the system
at time $t$ and the parameter $g$ depends on $(\upr,\eps,D)$ through an expression given in the \SOM.
For large enough $\tobs$, bulk properties
of the $s$-ensemble and symmetrized $s$-ensemble are the same.
In particular, the mean activity density within the 
symmetrized $s$-ensemble is
\begin{equation}
k(s) \equiv \frac{1}{N\tobs} \langle K \rangle_{s,\mathrm{sym}}.
\end{equation}
which depends implicitly on $N$ and $\tobs$ as well as the parameters of the model.
However, as $\tobs\to\infty$, then the activity densities in symmetrized and
original (unsymmetrized) $s$-ensembles approach the same limit, which is
$-N^{-1}\mathrm{d}\psi(s)/\mathrm{d}s$.

We show both first-order and second-order dynamical phase transitions
in Fig.~\ref{fig:phaseD}.  For fixed $\upr$, we show
a space-time phase diagram in the $(s,\eps)$ plane, with $D$ being
adjusted as in Equ.~\ref{equ:freef}.  The overall structure mirrors
that of Fig.~\ref{fig:phase_sketch}(b), but we are now working at fixed $\upr$
and varying $\eps$, whereas both of these parameters were varying separately
with temperature in Fig.~\ref{fig:phase_sketch}(b), as discussed above. 
 The condition of Equ.~\ref{equ:dual} is
shown as a dashed line, together with a solid first-order phase boundary that follows this line 
from the origin to the (known) position of the critical point.
We also show histograms of the activity $K$ obtained from $s$-ensembles
at the coexistence point $s=s^*$,
and the behavior of $k(s)$ as $s$ is varied.  We emphasize the analogy between the data
of this figure and the behavior of
a ferromagnetic model at phase coexistence: the bimodal distribution $P(K)$ is analogous to the
bimodal distribution of the magnetization at zero magnetic field, while the sharp change in $k(s)$ as $s$
is increased is analogous to the jump in the magnetization as the field is varied through zero.

Pursuing this analogy with ferromagnetic phase transitions,
we expect
a true jump discontinuity in $k(s)$ at $s=s^*$ only in the limit $N,\tobs\to\infty$.
(A discussion
of the order of the limits of $N$ and $\tobs$ is given in the \SOM.)
We show a finite-size scaling analysis of $k(s)$ in Fig.~\ref{fig:scaling}.
The data in Fig.~\ref{fig:scaling}(a) show an increasingly sharp jump in $k(s)$
as $N$ and $\tobs$ are increased.  On the other hand, Fig.~\ref{fig:scaling}(b)
shows the behaviour found at the critical point, while the inset shows how the maximal
susceptibility
$\chi^*=-\mathrm{d}k(s)/\mathrm{d}s|_{s=s^*}$ scales with the system size.
For $\eps<\eps_c$,
the data in the inset are consistent with the behavior at a first-order phase transition: 
$\chi^* =(\Delta k)^2 N\tobs/2$, with $\Delta k$ being the size of the jump in $k(s)$ 
at the phase transition.  For $\eps=\eps_c$, the dependence of $\chi^*$ on
$N$ and $\tobs$ is weaker, consistent with a second-order phase transition; for
$\eps>\eps_c$, the susceptibility $\chi^*$ is independent of the system size.

The numerical results of this section support our assertion 
that the phase diagram sketched in Fig.~\ref{fig:phase_sketch}(b) does indeed apply 
to the space-time phases of the sFA model, and we have introduced an analogy with phase transitions in ferromagnetic
systems.   
We now develop this analogy into a rigorous mapping, whose implications we will discuss in the final section.

\section{Theoretical analysis}

We write the master equation of the sFA model compactly, as 
\begin{equation}
\frac{\partial}{\partial t} |P\rangle = \WW |P\rangle
\label{equ:master}
\end{equation}
where $|P\rangle$ represents a probability distribution over the configurations
of the system, and $\WW$ is a linear operator whose matrix elements are the
transition rates of the sFA model. The $s$-ensemble may then
be studied by defining an operator $\WW(s)$ such that $\WW(0)=\WW$, and
the largest eigenvalue of $\WW(s)$ is the dynamical free energy $\psi(s)$. 
Details are given in the \SOM.
Since the sFA model obeys detailed balance, 
one may write
$\WW(s) = \ee^{-\EE/2T} \HH(s) \ee^{\EE/2T}$, where $\EE$
is an energy operator, 
and $\HH(s)$ is a symmetric (Hermitian) operator with the same eigenvalues as $\WW(s)$.

These observations allow properties of large deviations in space-time to by obtained  
from ground state properties of a quantum many-body system with Hamiltonian $-\mathbb{H}(s)$.  
Such mappings between stochastic classical systems and deterministic quantum ones are 
well-established~\cite{Doi-Peliti}, and a recent review~\cite{stinch} covers the
relevant cases for this article.
As well as the mathematical mapping, there is also a useful physical analogy:  the phase boundaries shown in Fig.~\ref{fig:phase_sketch} are points at which the dynamical free energy  $\psi(s)$ has a non-analytic dependence on $s$ and on the parameters of the model. In the quantum systems, such singularities are quantum phase transitions (QPTs) [17], which have been studied extensively.

\subsection{Mapping to quantum and classical spin systems}

As discussed in the \SOM,
we use a spin-half representation of the binary spins of the sFA 
model, following~\cite{FA-AA,stinch}.
The result is 
\begin{equation}
 -\HH(s) = NC - \sum_i (h_x \sig^x_i - h_z \sig^z_i) 
  - 
   \sum_{\langle ij\rangle} \sum_{\mu\nu} 
   \sig^\mu_i M^{\mu\nu} \sig^\nu_j
\label{equ:defH}  
\end{equation}
where $\mu,\nu\in\{x,y,z\}$ and the $\sigma^\mu_i$ are
Pauli matrices associated with site $i$, and $C=[D+(1+\upr)(1+\eps)]d/2$.  The scalars $h_x$ and $h_z$
and the coupling matrix $M$ depend on the sFA parameters $(\upr,\eps,D,s)$,
through expressions that are given in the \SOM.
The sum over $\langle ij\rangle$ runs over pairs of nearest neighbor
sites on a $d$-dimensional lattice.

This operator may be analysed in a mean-field approximation, following Ref.~\cite{Garrahan-Fred-PRL}.  
We replace operators in Eq.~\ref{equ:defH}
by numbers: $\sig^z_i\to2\rho-1$ and $\sig^x_i\to2\sqrt{\rho}$, where $\rho\ll1 $ is the mean
density of excitations.  We also take
$D=0$ and $\upr\ll1$ although these conditions may be relaxed at the expense of some algebra.
The result is a  
space-time Landau free energy: 
\begin{equation}
{\cal F}(\rho) = dN ( 2\rho + \eps ) ( \rho + \upr - 2 \ee^{-s} \sqrt{\rho \upr}).
\end{equation}
We have $\psi(s)\geq-\min_\rho {\cal F}(\rho)$~\cite{Garrahan-Fred-PRL,Garrahan-Fred-JPA} and
we use this bound to estimate $\psi(s)$.  The function $\FF(\rho)$
is quartic in $\sqrt{\rho}$, and may have either one or two minima.  At this mean-field
level, the point where the
single minimum bifurcates into two is the critical point, while cases where $\FF(\rho)$ has
two degenerate minima correspond to space-time phase coexistence.  For fixed $\upr$, the mean-field
estimate of the position of the critical point is $(\eps,s)=(2\upr/5,\log(\sqrt{5}/2))$, at
which $\FF(\rho) = 2dN[ (\sqrt{\rho} - \sqrt{\upr/5})^4 + (2\gamma/5)^2 ]$.

To move beyond this mean-field level, we
interpret $[-\HH(s)]$ as a quantum Hamiltonian and diagonalize
the matrix $M$.  That is, we let $\HH'=R^{-1}\HH R$ where $R$ is a 
uniform rotation of the spins (see \SOM), 
so that
\begin{equation}
-\HH'(s)
  = NC - \sum_i (B \sig^x_i - h \sig^z_i) 
 - \sum_{\langle ij\rangle} \sum_\mu J_\mu \sig^\mu_i \sig^\mu_{j},
 \label{equ:H_sym}
\end{equation}
with $(B,h,J_{x,y,z})$ being new constants that depend on $(\upr,D,\eps,s)$ through expressions given
in the \SOM.  For the quantum spin system described by $[-\HH'(s)]$, $h$ and $B$ are magnetic field
terms: we have $h>0$ but $B$ may have either sign.  The most interesting behavior of the
quantum system occurs when $B=0$, in which case the field $h$ tends to align
the spins along the $-\sig^z$ direction, while the ferromagnetic coupling $J_x$ 
promotes ferromagnetic ordering along $\pm\sig^x$ directions. 
For $B=0$ and small $h/J_x$ there is a single ground state
with $\langle \sig^x\rangle=0$.  However,
for $B=0$ and large $J_x/h$ there are two
degenerate ground states, with the symmetry of $\HH'$ under $\sig^x\to-\sig^x$ being
spontaneously broken.  These two regimes are separated by a quantum phase transition.

There is a standard exact mapping between quantum spin systems in $d$ dimensions and
classical spin systems in $(d+1)$ dimensions~\cite{Sachdev}. One takes a small time increment $\delta t$ and
considers
the operators $\ee^{\HH(s)\delta t}$ and $\ee^{\HH'(s)\delta t}$ as transfer matrices 
that generate ensembles of configurations for $(d+1)$-dimensional
Ising systems.  We denote the ensembles
defined by these two transfer matrices as
$\Sc$ and $\Sc'$ respectively, with details given in the \SOM.  
The ensembles are defined for classical Ising systems
on anisotropic $2d$ lattices, but we concentrate
here on symmetries and universal properties of the models, which do not
depend on its underlying lattice.  The weights of configurations in ensemble $S$
are identical to those of corresponding trajectories of the sFA model in the $s$-ensemble,
with the time axis in the sFA model being interpreted as the $(d+1)$th spatial
axis in the classical Ising system.

\subsection{Consequences for the sFA model}

One consequence of the mapping between $\WW(s)$ and $\HH'(s)$ is that the condition for space-time phase coexistence in the sFA model, Equ.~\ref{equ:dual}, is simply the condition that $B = 0$ in the quantum spin model, Equ.~\ref{equ:H_sym}.
This means that $\HH'(s)$ is unchanged
by the global transformation $\sig^x_i\to-\sig^x_i$, and 
the classical ensemble $\Sc'$
is also symmetric under global spin inversion, except for possible boundary effects
that are discussed below.  This symmetry of $\HH'$ corresponds to a symmetry
transformation of the sFA model that relates active and inactive phases, and is
used to derive Equs.~\ref{equ:dual} and~\ref{equ:ens-sym}.  
Thus, the dashed line 
shown in Fig.~\ref{fig:phaseD}(a) for the sFA model corresponds to a zero-field
condition for the quantum and classical magnetic systems.
When distinct active and inactive phases exist in the 
sFA model, they correspond to ferromagnetic phases in the quantum spin system
$\HH'(s)$, and to classical ferromagnetic phases in ensemble $\Sc'$.  

As usual in ferromagnetic systems, the coexistence line between the
ferromagnetic phases ends at a critical point.
The ensemble $\Sc'$ has the symmetry properties of an 
Ising model in $(d+1)$ dimensions.
Thus, the finite-temperature critical point shown in
Fig.~\ref{fig:phase_sketch}(b) is in the universality class
of a $(d+1)$-dimensional Ising model.  For this
critical point, the (upper) critical dimension
of the sFA model is $d_c=3$, while the dynamical exponent that sets the relative
scaling of space and time is $z=1$ in all dimensions~\cite{Sachdev}.
This is in contrast with the zero-temperature critical points in Fig.~\ref{fig:phase_sketch}, 
where $z=2$ and $d_c=2$~\cite{FA-AA}.

The mapping to a classical ensemble $\Sc'$ also motivates our definition of
the symmetrized $s$-ensemble.  As we discuss in the \SOM, 
the $s$-ensemble of Equ.~\ref{equ:s-vanilla} corresponds to an ensemble $\Sc'$
in which boundary conditions are biased towards one of the ferromagnetic phases.
However, the 
symmetrized $s$-ensemble 
defined in Equ.~\ref{equ:ens-sym} corresponds to an ensemble
$\Sc'$ that is invariant under global spin inversion. 
Thus, the symmetrized $s$-ensemble removes 
biases towards active or inactive phases in the sFA model by ensuring that 
ensemble $\Sc'$ has no symmetry-breaking bias.
This property enabled the accurate
finite-size scaling analysis shown in Fig.~\ref{fig:scaling}.
Our choice of the parameter $D$ given in Equ.~\ref{equ:freef}  is also motivated by
properties of $\HH'(s)$: 
we show in the \SOM~that if both Equ.~\ref{equ:dual}
and Equ.~\ref{equ:freef} are satisfied, then $B=J_z=0$. In one dimension,
$\HH'(s)$ may then be diagonalized by a Jordan-Wigner transformation~\cite{stinch},
allowing us to locate the critical point in Fig.~\ref{fig:phaseD}.

Finally, it is useful to generalize concepts of equilibrium thermodynamic phases 
to those of non-equilibrium space-time 
phases in the $s$-ensemble.  
For example, if ensemble $\Sc'$ contains coexisting phases,
one may calculate the surface tension $\Gamma$ between them.  In the
sFA model, $\Gamma$ gives the value of a space-time surface tension whose interpretation will be given in the
final section.  Away from phase coexistence ($B\neq0$), ensemble $\Sc'$ is dominated by a single
phase, but one may still investigate
metastable phases with opposite magnetization.  In particular, the free
energy difference between phases and the spinodal limit of stability of the metastable phase may
be calculated approximately.
Variational and mean-field
methods for estimating such space-time free energy differences and spinodals in 
$s$-ensembles for KCMs
were discussed in Ref.~\cite{Garrahan-Fred-JPA} and we summarize them in the \SOM. 

\section{Implications of these results for the glass transition}

The critical point we have uncovered appears as kinetic constraints are softened, and its discovery has implications for natural systems even while experimental methods to directly access the $s$-ensemble are not yet known.  To explain this, 
we return to our analogy between sFA and ferromagnetic systems. 

Consider an ordered ferromagnetic system, below its critical temperature and in a small magnetic field.  Within the
ordered equilibrium state (which has positive magnetization), the dominant fluctuations are small domains of the 
minority phase (with negative magnetization). The probability of observing such a domain depends on the surface tension and the free energy difference between majority and minority phases.

In the models that we have considered, dynamical heterogeneities
occur by a similar mechanism: the dominant fluctuations in the (active) supercooled liquid state come from domains of the inactive space-time phase. In that case, the probability of observing inactive behavior in a region of space-time is set by several factors: the spatio-temporal extent of the region, and the surface tension and free energy difference between dynamical phases. In the unbiased equilibrium dynamics of the system, one expects the dominant contributions to this probability to take the form~\cite{spacetime-jcp}
\begin{equation}
P(\ell^d,\tau) \propto \exp(-\Gamma_1\tau - \Gamma_2\ell^d - \Delta \psi \ell^d \tau)
\label{equ:bubbles}
\end{equation}
where $\ell^d$ and $\tau$ are the spatial and temporal extents of the inactive domain,
$\Gamma_{1,2}$ are surface tensions,
and $\Delta\psi$ is the difference in free energy between the space-time phases, evaluated at $s=0$.

We have shown~\cite{spacetime-jcp,Garrahan-Fred-PRL} that KCMs (with $\eps=0=s^*$)
lie naturally at phase coexistence so that $\Delta\psi=0$.  However, for models with
$s^*>0$ (including the sFA model) we expect $\Delta\psi>0$.  In either case,
the probability of observing large dynamical heterogeneities in the
system is set by the space-time surface tensions and free energies.  
Of course, in defining the distribution $P(\ell^d,\tau)$,
one assumes the existence of two space-time phases, which is strictly valid
only at phase coexistence.  However, one may use a mean-field spinodal condition
for the classical spin model to estimate
whether the minority (inactive) space-time phase is sufficiently stable to form fluctuating
domains within the stable active state, as in the magnetic case.

Based on these arguments, we now return again to
Fig.~\ref{fig:phase_sketch}(b).  For high temperatures in the sFA model,
there is only a single phase, which we identify with a simple liquid.  As the
system is cooled, a new inactive phase comes into existence at a critical
point, at which $s>0$.  
Whether the inactive phase has observable consequences in the liquid
depends on its stability at equilibrium ($s=0$), 
and on the free energy difference and surface tension 
between active and inactive phases. 
For the sFA model, these factors
be estimated via mean-field arguments, as discussed above.
Within the theoretical picture presented here, the stability of the inactive space-time 
phase and the parameters $(\Gamma_1,\Gamma_2,\Delta\psi)$ are the key quantities that 
determine the nature of the dynamical heterogeneities in supercooled liquids.

Finally, the phase diagram of Fig. \ref{fig:phase_sketch}b connects our work to different theoretical scenarios for the glass transition.  First, if cooling a supercooled liquid 
is analogous to reducing both $\upr$ and $\epsilon$ in the sFA model, then both the dynamical free energy difference and the surface tension between the phases vanish as $T \rightarrow 0$. This corresponds to a zero-temperature ideal glass transition for the liquid~\cite{Whitelam}. Such transitions are accompanied by increasing dynamical heterogeneity since the probability of large inactive space-time regions increases, according to Equ.~\ref{equ:bubbles}.

Second, if molecular liquids support a finite-temperature ideal glass transition along the lines of the thermodynamic
glass transition in spin glasses~\cite{RFOT}, one expects the dynamical free energy difference and surface tension to vanish at that point. (For a mean-field analysis of that situation in the $s$-ensemble, see Ref.~\cite{s-ROM}.)

Third, for the active-inactive phase boundary of Fig.~\ref{fig:phase_sketch}(b) to end at a critical point as it does for the sFA model, 
the two phases must have the same symmetry properties. (Critical end-points for phase boundaries separating crystalline and liquid 
phases are forbidden for this reason.)  In a molecular system, it is not clear \emph{a priori} whether the inactive phase should be a true amorphous solid that spontaneously breaks translational symmetry, or a yet-to-be-observed liquid phase with an extraordinarily large but finite relaxation time. In the former case, the critical end-point shown in Fig.~\ref{fig:phase_sketch}(b) cannot occur, and any active-inactive phase boundary must separate the $(s,T)$ plane into distinct regions.  However, the latter possibility -- a liquid-liquid transition that is relevant for the glass transition~\cite{Tarjus-Kiv,Tanaka,Stanley} -- can be consistent with the critical end-point discussed here, provided that the liquid-liquid transition is a non-equilibrium transition.

\begin{acknowledgments}
We thank Fred van Wijland and Peter Sollich for helpful discussions.
This research has been funded by the US National Science Foundation (with a fellowship for YSE), 
by the US Department of Energy (with support of DC, Contract No. DE-AC02-05CH11231), 
and, in its early stages, by the US Office of Naval Research (with support of RLJ, Contract No. N00014-07-10689)

\end{acknowledgments}

\section{Appendix}


\renewcommand{\theequation}{A\arabic{equation}}
\setcounter{equation}{0}

\newcommand{\lastref}{S2}

\newcommand{\citemain}[1]{~[#1]}

\newcommand{\DoiP}{25}
\newcommand{\JMS}{20}
\newcommand{\TPS}{24}
\newcommand{\fred}{23}
\newcommand{\jpfred}{9,10}
\newcommand{\sachdev}{18}
\newcommand{\stinch}{26}
\newcommand{\eqdual}{3}
\newcommand{\eqsym}{5}
\newcommand{\eqff}{4}
\newcommand{\figphase}{3}

\newcommand{\C}{\mathcal{C}}
\newcommand{\phibar}{\overline\phi}

\parindent0pt
\parskip6pt




\subsection{Numerical Methods}
\label{sec:num}

The sFA model is simulated by a continuous time Monte Carlo
method~\cite{NB-book}.  
To efficiently sample trajectories within the $s$-ensemble we use 
transition path sampling (TPS)\citemain{\TPS} .  Starting from 
an equilibrated initial condition, we simulate a trajectory
of length $t_\mathrm{obs}$, storing the configuration of the
system at a set of equally-spaced times.
We then generate new
trajectories using ``half-shooting'' and
``shifting'' TPS moves\citemain{\TPS}, starting from the stored configurations.

To sample the equilibrium ensemble of trajectories, we 
accept all trajectories generated in this way: 
this corresponds to an unbiased
random walk in trajectory space.  To sample
the symmetrised $s$-ensemble, we use a Metropolis-like criterion for
acceptance or rejection of the trial TPS move.  For each trajectory, we calculate 
the quantity 
\begin{equation}
{\cal E} = sK - g[\mathcal{N}(0)+\mathcal{N}(\tobs)]
\end{equation}
where $K$ is the number of configuration changes in the trajectory,
 $\mathcal{N}(t)=\sum_{i=1}^N n_i(t)$ is the number of excited sites in the system at time $t$, and the
fields $s$ and $g$ depend on the ensemble being sampled, as described in the main text.
[An expression for $g$ is given in (\ref{equ:def_g}) below.]
We then
 compare
the value of  $\cal E$ for the original (old) trajectory
and for the new trajectory generated by TPS.  We accept the new
(trial) trajectory with a probability
$P_\mathrm{acc}^0=\mathrm{min}\left\{1,\mathrm{exp}
[{\cal E}_\mathrm{old}-{\cal E}_\mathrm{trial})]\right\}$.  
Generalising the results of Refs.~[\TPS] and \cite{tps_stoch} it can be shown that these rules
respect detailed balance in the space of trajectories, according
to the distribution $P_s[x(t)]$.  Thus, after repeating many such moves,
the algorithm generates trajectories according to this distribution.



\subsection{Definitions of $\WW(s)$, $\HH(s)$ and $\HH'(s)$}

The master equation of the sFA model takes the standard form
\begin{equation}
\partial_t P(\C,t) = -r(\C) P(\C,t) + \sum_{\C'} W(\C'\to\C) P(\C',t)
\end{equation}
where
$P(\C,t)$ is the probability that the system is in
some configuration $\C$ at time $t$, the
$W(\C'\to\C)$ are the rates for transitions between configurations, 
and $r(\C)=\sum_{\C'} W(\C\to\C')$

We use a spin-half representation
of the master equation of the sFA model\citemain{\stinch}.  A configuration of
the system is specified by the spin variables $n_i$ with $i=1\dots N$.
We represent the $\{n_i\}$ by $N$ quantum spins, and we denote the state
with all spins down ($n_i=0$) by $|\Omega\rangle$.
Then, if $\sig_i^{x,y,z}$ are Pauli matrices associated with the sites, 
and $\sig^\pm_i=\frac12(\sig^x_i\pm\sig^y_i)$ as usual, then $\sig^-_i|\Omega\rangle=0$
by construction of $|\Omega\rangle$, while
a configuration of the sFA model is represented by
\begin{equation}
|\{n_i\}\rangle = \prod_{i=1}^N (\sig^+_i)^{n_i} |\Omega\rangle.
\end{equation}

We construct a ket state
\begin{equation}
|P(t)\rangle = \sum_\C P(\C,t) |\C\rangle,
\end{equation}
where the sum runs over all configurations of the system.
Then, the master equation is
\begin{equation}
\frac{\partial}{\partial t}|P(t)\rangle = \WW |P(t)\rangle
\end{equation}
with
\begin{multline}
\WW = \sum_{\langle ij\rangle} (\hat n_j + \eps/2) [ (1-\sig_i^+) \sig_i^- + \upr(1-\sig_i^-) \sig_i^+ ]
 \\
   + D [\sig^+_i \sig^-_j -  (1-\hat n_i) \hat n_j] + (i\leftrightarrow j) 
\end{multline}
where the sum is over (distinct) pairs of nearest neighbors. 
We define
 $\hat n_j\equiv\sig^+_j \sig^-_j$, and the notation $(i\leftrightarrow j)$ indicates that the entire
summand is to be symmetrised between sites $i$ and $j$.

\subsubsection{Operator representations of biased ensembles}

To investigate the model within the $s$-ensemble, we follow Ref.~[\fred] in writing $P(\C,K,t)$ for 
the probability of being in configuration $\C$ at time $t$, having accumulated $K$
configuration changes between times $0$ and $t$.  Then, we write
$P(\C,s,t)=\sum_\C P(\C,K,t) \ee^{-sK}$ and consider the equation of motion for $|P(s,t)\rangle
=\sum_\C P(\C,s,t) |\C\rangle$, which is
\begin{equation}
\frac{\partial}{\partial t}|P(s,t)\rangle = \WW(s) |P(s,t)\rangle
\end{equation}
with
\begin{multline}
\WW(s) = \sum_{\langle ij\rangle}
(\hat n_j + \eps/2) [ (\ee^{-s}-\sig_i^+) \sig_i^- 
 + \upr(\ee^{-s}-\sig_i^-) \sig_i^+ ] \\
 + D  [\ee^{-s} \sig^+_i \sig^-_j - (1-\hat n_i) \hat n_j] 
+ (i\leftrightarrow j).
\end{multline}

The dynamics of the sFA model respect detailed balance, with an energy function $E=J\sum_i n_i$ and
$\upr=\ee^{-J/T}$, 
so we define an energy operator $\EE=J\sum_i \hat n_i$.  Then, defining
$$
\HH(s) \equiv \ee^{\EE/2T} \WW(s) \ee^{-\EE/2T}
$$
it may be verified that
\begin{multline}
\HH(s)
 = \sum_{\langle ij\rangle}
 (\hat n_j + \eps/2) [ (\sqrt{\upr}\ee^{-s}-\sig_i^+) \sig_i^- 
 + (\sqrt\upr\ee^{-s}-\upr\sig_i^-) \sig_i^+ ] 
\\
 + D [\ee^{-s} \sig^+_i \sig^-_j -  (1-\hat n_i) \hat n_j] + (i\leftrightarrow j)
\end{multline}
is a Hermitian (symmetric) operator, i.e. $\langle \C|\HH(s)|\C'\rangle = \langle \C'|\HH(s)|\C\rangle$.

Writing $\HH(s)$ in terms of the $\sig^{x,y,z}$, we recover 
\begin{equation}
\HH(s) = -NC + \sum_i ( h_x \sig_i^x - h_z \sig_i^z) + \sum_{\langle ij\rangle} \sum_{\mu\nu}
\sig_i^\mu M^{\mu\nu} \sig_j^\mu,
\end{equation}
with 
\begin{eqnarray}
h_x&=&dz(1+\eps)\sqrt{\lambda},
\nonumber \\ 
h_z&=&d[2+\eps-\upr\eps]/2,
\end{eqnarray}
and
\begin{equation}
M = \frac12 \left( \begin{array}{ccc} zD & 0 & z\sqrt\upr \\ 0 & zD & 0 \\ z\sqrt\upr & 0 & D + \lambda -1 
\end{array} \right).
\end{equation}
We use the shorthand notation $z=\ee^{-s}$ for convenience.

Finally, we make a rotation of the spins, letting $R(\alpha)=\ee^{\ii\alpha\sum_j \sig^y_j/2}$
so that 
\begin{equation}
R(-\alpha) \left(\begin{array}{c}\sig^x_i \\ \sig^y_i \\ \sig^z_i\end{array}\right) R(\alpha)
 = \left(\begin{array}{c}\sig^x_i\cos\alpha - \sig^z_i\sin\alpha \\
 \sig^y_i \\ \sig^z_i\cos\alpha+\sig^x_i\sin\alpha \end{array} \right)
\end{equation}
We choose 
\begin{equation}\tan2\alpha=\frac{2z\sqrt\upr}{1-\lambda-D(1-z)}, \label{equ:t2a}\end{equation} 
so as to diagonalise $M$.  
That is
\begin{eqnarray}
\HH'(s) &\equiv& R(-\alpha) \HH(s) R(\alpha)
\nonumber \\
 &=&
 -NC + \sum_i ( B \sig_i^x - h \sig_i^z) + \sum_{\langle ij\rangle} \sum_{\mu}
J_\mu \sig_i^\mu \sig_j^\mu \nonumber \\
\end{eqnarray}
where
\begin{eqnarray}
B&=& h_x\cos\alpha - h_z\sin\alpha,
\nonumber \\ 
h &=& h_z\cos\alpha + h_x\sin\alpha,
\end{eqnarray}
and $J_{x,y,z}$ are the eigenvalues of the matrix $M$.

\subsubsection{Interpretation of $\HH(s)$ and $\HH'(s)$ as transfer matrices, and ensembles $S$ and $S'$}

In the main text, we discussed how singularities in the ground state energy of $[-\HH'(s)]$ may be interpreted
as quantum phase transitions.  Alternatively, one may interpret $\HH(s)$ as a transfer matrix for a classical
spin system in $(d+1)$ dimensions.  This mapping between quantum and classical systems is now standard~[\sachdev],
the only subtlety being that the classical system has discrete co-ordinates along the spatial axes of the relevant
quantum system, but the extra $(d+1)$th dimension in the classical system is a continuous co-ordinate.
This leads to a direct analogy between trajectories of the sFA model, and configurations of the classical
spin system in $(d+1)$ dimensions, on this slightly unusual anisotropic lattice.

The most natural approach is to discretise the time axis using a small time $\delta t$.  Then, one may interpret
the sequence of $d$-dimensional configurations at time $0,\delta t, 2\delta t, \dots$ in the sFA model
as `planes' in a $(d+1)$-dimensional classical spin model.  This may be achieved by taking $\ee^{\HH(s)\delta t}$ 
as a classical transfer matrix. That is, $\langle \C | \ee^{\HH(s)\delta t} |\C'\rangle$ is proportional to the
the probability that the final plane of a system is in configuration $\C$,
 given that its penultimate plane is in configuration $\C'$.  For constructing the ensemble
$S$, one uses the $\sig^z$ components of the spins in $\C$ to give the states of the classical Ising
spins, as in the sFA model.

However, for the ensemble $S'$ formed from $\ee^{\HH'(s)\delta t}$, we make a different choice.  We associate
up (down) spins in $S'$ with spins in $|\C\rangle$ that are aligned along the positive (negative) $\sig^x$ direction.
This ensures that $S'$ has the appropriate symmetries when the $\sig^x$ component of the spins in $\HH'(s)$
are inverted.  This means that the configurations of $S'$ do not have a 
straightforward 
relation with the trajectories
of the sFA model.  However, one may always relate expectation values in the two ensembles by writing
them as Dirac brackets, as discussed in Sec.~\ref{subsec:sym}, below.

Finally, it is also important to consider the boundary conditions associated with ensemble $S'$. 
For the $d$ spatial dimensions of the sFA model,
we take periodic boundaries, corresponding to periodic boundaries in $S'$.  However, for the $(d+1)$th
dimension in $S'$, the boundary conditions depend on the initial and final conditions
for the $s$-ensemble.  These conditions are specified in turn by the initial condition of the unbiased ($s=0$,
equilibrium) average $\langle \cdot \rangle_0$ used in the definition of the $s$-ensemble.  Consequences of
these boundary conditions are discussed in Sec.~\ref{subsec:sym}, below

\subsubsection{Order of limits of $N$ and $\tobs$}

As discussed in the main text, we consider trajectories where $\tobs$ is very long,
and we also consider systems where $N$ is large.  For example, the sFA model
in one dimension maps to the two-dimensional Ising model, and in that case we must take a 
limit of large system size both parallel and perpendicular to the transfer direction
in order to observe any phase transition.
%
When considering systems evolving in time, it is usual to take the limit of large system size $N$
before any limit of large time $\tobs$.  However, our theoretical analyses based on the space-time free
energy $\psi(s)$ implicitly assume a limit of large-$\tobs$ before large-$N$.  Based on physical
considerations, we expect these limits to commute but we have not verified this in our analysis.

\subsection{Symmetries of $\HH'(s)$}

\subsubsection{Necessary condition for phase coexistence}

An important special case for the sFA model occurs when $B=0$, since $\HH'(s)$ is then invariant under $\sig_i^x\to-\sig^x_i$.
As in the main text, we interpret $[-\HH'(s)]$ as the Hamiltonian for a quantum spin system, and
this symmetry may be spontaneously broken in the ground state, if $J_x/h$ is sufficiently large.  
The condition $B=0$ occurs for $\tan\alpha=2\sqrt{\lambda}z(1+\eps)/(2+\eps-\eps\upr)$.  
Combining this condition for $\alpha$
with (\ref{equ:t2a}), one arrives at the condition for $B=0$:
\begin{equation}
\frac{1+\upr}{1+\eps} = \sqrt{ [1-\lambda-D(1-z)]^2 + 4z^2\upr } - D(1-z).
\end{equation}
This is consistent with Equ.~(\eqdual) of the main text.


\subsubsection{Construction of symmetrized $s$-ensemble}
\label{subsec:sym}

We now motivate our definition of the symmetrized $s$-ensemble, as a tool for
accurate characterization of space-time phase coexistence.  For convenience, we consider
the behavior of a one-time observable $F(t)$ in the $s$-ensemble.  If $F(t)$ has
different expectation values in the two phases, one expects it to cross over sharply
between these two values, as $s$ is tuned through its coexistence value $s^*$.  
By casting the expectation of $F(t)$ as an observable in the thermodynamic ensemble $S'$, 
we now explain why the symmetrized $s$-ensemble is superior to the $s$-ensemble for characterizing
the behavior of $F(t)$ near space-time phase coexistence.

In the main text, we define $s$-ensembles through their expectation values.  
The expectation value of $F(t)$ may be written in terms of
Dirac brackets
\begin{equation}
\langle F(t) \rangle_s =  \frac{ \langle - | \ee^{\WW(s) (\tobs-t)} \hat F \ee^{\WW(s)t} | \mathrm{eq} \rangle }
                              { \langle - | \ee^{\WW(s) \tobs} | \mathrm{eq} \rangle }
\label{equ:dirac}
\end{equation}
where $\hat F$ is the operator corresponding to the observable $F$,
$\langle - | = \langle\Omega|R(-\pi/2)$ is a projection state, and
$|\mathrm{eq}\rangle= R(2\chi)|\Omega\rangle$ is the equilibrium state, with $\tan\chi=\lambda$.

If we then use the similarity transform 
$$\HH'(s)=R(-\alpha)\ee^{\EE/2T}\WW(s)\ee^{-\EE/2T}R(\alpha),$$
we have
\begin{equation}
\langle F(t) \rangle_s =  \frac{ \langle \Psi | \ee^{\HH'(s) (\tobs-t)} \hat F' \ee^{\HH'(s)t} | \Psi \rangle }
                              { \langle \Psi | \ee^{\HH'(s) \tobs} | \Psi \rangle }
\label{equ:FS'}                              
\end{equation}
with $|\Psi\rangle=R(2\theta-\alpha)|\Omega\rangle$ and $\langle \Psi |$ its Hermitian conjugate, with $\tan\theta=\sqrt\upr$.
%
%
This equation may also be interpreted as a transfer matrix representation of an expectation value in ensemble
$S'$, which may be seen by writing the numerator of  (\ref{equ:FS'})
as 
\begin{multline}
\sum_{\C_0\cdots \C_M} h(\C_M) \left[\prod_{i=m+1}^{M-1} U(\C_{i+1},\C_i)\right] U(\C_{m+1},\C_m) \\ \times F'(\C_{m+1},\C_m) 
\left[\prod_{i=0}^{m-1} U(\C_{i+1},\C_{i})\right] h(\C_0)
\end{multline}
 where the $\C_i$ are configurations of the planes of the system,
which are to be summed over, with $M=\tobs/\delta t$ and $m=t/\delta t$.  Here, $h(\C)$, $U(\C,\C')$ and $\hat F'(\C,\C')$ are matrix elements
of $|\Psi\rangle$, $\ee^{\HH'(s)\delta t}$ and $\hat F'$, where 
$\hat F'=R(-\alpha)\ee^{\EE/2T}\hat F \ee^{-\EE/2T} R(\alpha)$ corresponds
to a new observable to be measured in ensemble $S'$.

If space-time phase coexistence occurs
in the sFA model, then the ensemble $S'$ is also at phase coexistence. For this to occur,
 $\HH'(s)$ should be invariant under inversion of $\sig^x$.  However, for ensemble $S'$ to
 be invariant under a global spin flip, one requires not just that the transfer matrix be
 invariant, but also that the boundary conditions along the transfer direction are unbiased
 between the coexisting phases.
Within the $s$-ensemble, there is a finite boundary bias.  This is apparent from Equ.~(\ref{equ:FS'})
since the state $|\Psi\rangle$ is not invariant under inversion of $\sig^x$ (in general, $\theta\neq2\alpha$). 
 This leads to a predominance 
of one phase over the other near the boundaries in $S'$.

To accurately characterize phase coexistence in the classical system, one may replace
$|\Psi\rangle$ in (\ref{equ:FS'}) by a new state vector that is symmetric between the two phases: the natural
choice is to replace $|\Psi\rangle$ with $|\Omega\rangle$ 
(and similarly $\langle\Psi|$ by $\langle\Omega|$).  
Making this replacement, and transforming back to the sFA representation, this symmetrized
matrix element takes the form
\begin{equation}
\langle F(t) \rangle_{s,\mathrm{sym}} =  \frac{ \langle - | \ee^{g \sum_i \hat n_i}
 \ee^{\WW(s) (\tobs-t)} \hat F \ee^{\WW(s)t} \ee^{g \sum_i \hat n_i} | \mathrm{eq} \rangle }
                              { \langle - | \ee^{g \sum_i \hat n_i} \ee^{\WW(s) \tobs} \ee^{g \sum_i \hat n_i} | \mathrm{eq} \rangle }
\end{equation}
with 
\begin{equation}
\ee^g = \tan(\alpha/2)/\sqrt{\lambda}.
\label{equ:def_g}
\end{equation}
which defines $g$.

It may be 
verified that this result is equivalent to Equ.~(\eqsym) of the main text, for the expectation
value of $F(t)$ in the symmetrized $s$-ensemble.  The generalization to more complex observables $A$
is trivial, requiring only a slightly heavier notation.  Thus, the symmetrized $s$-ensemble allows accurate
characterization of phase coexistence, which may be verified by showing that it corresponds to removal
of boundary biases in expectation values of an equivalent magnetic system.

\subsubsection{Conditions for free fermion solution, and analytic calculation of phase diagram}

A further special case occurs in $d=1$, when $B=0$ and $J_z=0$ together.  
In this case, the only couplings
in $\HH'(s)$ are $(h,J_x,J_y)$ and the model may be diagonalized by a Jordan-Wigner transformation\citemain{\stinch,\sachdev}.
Using these standard methods, one finds that spontaneous symmetry breaking occurs for $h<J_x+J_y$,
with criticality occurring for $h=J_x+J_y$.

By calculating the determininant of $M$,
it can be seen that $J_z=0$ if 
\begin{equation}D(D+\lambda-1)=z\lambda.
\end{equation}
  Solving for $D$ gives
Equ.~(\eqff) of the main text.  


\subsection{Mean-field approximation and construction of field theory}


As discussed in the main text, the space-time phase transitions of the sFA 
model in $d$-dimensions are closely related to symmetry-breaking
in Ising-like models in $(d+1)$ dimensions.  There are several ways to demonstrate this.  
One option is to generalzse the sFA model to allow sites to contain more than one excitation.
This allows the master equation to be written in a bosonic representation due to Doi and Peliti\citemain{\DoiP}.  
For small $\lambda$, it may be shown that the behavior of this generalized (`bosonic') sFA model
approaches that of the original model, via a large-$S$ expansion  (see, for example, the analysis of the 
FA model (with $\eps=0$) in Ref.~[\JMS]).  Further, even if $\lambda$ is not small, the universal (critical)
behavior of generalized and original sFA models are the same.  

Following the Doi-Peliti procedure for this generalized sFA model, the master operator within the $s$-ensemble is 
\begin{multline}
\WW_b(s) = \sum_{\langle ij\rangle} \left\{ (a^\dag_i a_i + \eps/2) [ (z-a^\dag_j) a_j + \upr (a^\dag_j - z) ]  + (i\leftrightarrow j) \right\} \\
 + D [ z(a^\dag_i a_j + a^\dag_j a_i) -(a^\dag_i a_i + a^\dag_j a_j) ]
\end{multline} 
where $a_i$ and $a^\dag_i$ are bosonic operators, so $[a_i,a_j^\dag]=\delta_{ij}$ as usual.
In treating this generalized model as an approximate representation of the sFA model, 
the parameters $(D,\upr,\eps)$ play the same role as in the original model.  Within
this new representation, the density
of excitations in the sFA model corresponds to the density of bosons, through the number
operators $a^\dag_i a_i$.

One may then use coherent state path integrals to represent Dirac brackets such as those
of Equ.~(\ref{equ:dirac}).  This is discussed, for example, in Ref.~[\JMS].  One arrives at
\begin{equation}
Z(s,\tobs) = \int \mathcal{D}(\phi,\phibar) \exp\left(-\int\!\mathrm{d}^dx\,\mathrm{d}t \,
L[\phi,\phibar]\right)
\end{equation}
where $\phi(x,t)$ and $\phibar(x,t)$ are complex conjugate fields, and
\begin{multline}
L[\phi,\phibar] = \phibar\frac{\partial \phi}{\partial t} - zD\ell_0^2 \phibar\nabla^2\phi +
   2d D(1-z)  \phibar\phi  \\ -  d (  2\phibar\phi + \eps\ell_0^{-d} ) [ z(\phi + \upr\phibar)\ell_0^{d/2} - (\upr + \phibar\phi\ell_0^d) ].
\end{multline}
%
%
Here, $\ell_0$ is the lattice spacing of the sFA model and we recall that the units of time have
been set by taking $\gamma=1$ in the original definition of the model.
The operators $a_i$ and $a_i^\dag$ have become
fields $\phi$ and $\phibar$, with the combination $\phi\phibar$ corresponding
to the density of excitations in the sFA model.

As discussed in Ref.~[\jpfred], one may either analyze the resulting field theory through
the saddle points of $L[\phi,\phibar]$ or by using a variational analysis that leads to a
free energy ${\cal F}(\phi)$, as in the main text.  The saddle points of $L[\phi,\phibar]$ give the properties
of space-time phases: free energies are obtained from the values of $L$ at while surface tensions
between phases may be estimated from inhomogeneous saddle points of the action.  The 
free energy ${\cal F}(\phi)$ may also be used to obtain spinodal conditions on the properties of
metastable phases~[\jpfred].

\end{article}

\end{document}